\documentstyle[11pt,epsfig]{article}

\newlength{\dinwidth}
\newlength{\dinmargin}
\def\squark{\widetilde q}

\def\sql{\squark_{L}}
\def\sqr{\squark_{R}}

\def\st{\widetilde{t}}

\def\sz1{{\widetilde{Z}}_{1}}

\def\msz1{m_{\sz1}}

\def\stl{\st_{1}}

\def\sth{\st_{2}}

\def\mst{m_{\st}}
\def\mstl{m_{\stl}}

\def\tht{\theta_{t}}

\begin{document}
~~~\\
\vspace{10mm}
\begin{flushright}
metro-hs-ph1 \\
AP-SU-00/02 \\
hep-ph/0012166 \\
Dec, 2000 
\end{flushright}
\begin{center}
  \begin{Large}
   \begin{bf}
Single scalar top production with polarized  beams in $ep$ collisions at HERA\\
   \end{bf}
  \end{Large}
  \vspace{5mm}
  \begin{large}
Shoichi Kitamura \\
  \end{large}
Tokyo Metropolitan University of Health Sciences, Tokyo 116-8551, Japan \\
kitamura@post.metro-hs.ac.jp\\
 \vspace{3mm}
 \begin{large}
    Tadashi Kon\\
  \end{large}
Faculty of Engineering,
Seikei University, Tokyo 180-8633, Japan \\
kon@apm.seikei.ac.jp\\
 \vspace{3mm}
 \begin{large}
    Tetsuro Kobayashi\\
  \end{large}
Faculty of Engineering,
Fukui Institute of Technology, Fukui 910-8505, Japan \\
koba@ge.seikei.ac.jp\\
  \vspace{5mm}
\end{center}
\vskip50pt
\begin{quotation}
\noindent
\begin{center}
{\bf Abstract}
\end{center}
From the point of view of the R-parity breaking supersymmetric model, we
propose a scalar top (stop) search with longitudinally polarized electron
($e^-_L$) and positron ($e^+_R$) beams which will soon be available at the
upgraded HERA.    Fully polarized $e^-$ or $e^+$ beams could produce the
stop two times as much as unpolarized beams, while they increase background
events due to the process of the standard model by  $\sim$30\% in comparison
with unpolarized ones.   We show that right-handed $e^+$ beams   at HERA is
efficient to produce the stop in the model. With 1 fb$^{-1}$ of integrated
luminosity  we estimate reach in the coupling constant $\lambda'_{131} $ for
masses of the stop in the range  160-400 GeV.   We can set a 95\%
confidence-level exclusion limit for  $\lambda'_{131} > $0.01-0.05  in  the
stop mass  range of 240-280 GeV if no singal of the stop is observed. We
also point out that $y(=Q^2/sx)$ distributions of $e^+$ coming from the stop
shows the different behavior from those of the standard model.
\end{quotation}
\vfill\eject
\section{\it Introduction}
In supersymmetric (SUSY) extension of the standard model(SM), a bosonic
superpartner (sfermion) is assigned to every fermionic SM particle and vice
versa\cite{nilles,haber}.  Helicity states of the SM quark $q_L$ and $q_R$
acquire SUSY partners $\sql$ and $\sqr $ , which are also the mass
eigenstates for the first two generations to a good approximation.   On the
other hand, a large mixing is expected between the left- and right-handed
states of scalar SUSY partner in the third generation because of the large
mass of  the fermionic SM particles\cite{hikasa}. This leads to a possible
large splitting between the mass states , which implies the existence of  a
lighter scalar top quark(stop,$\widetilde{t_1}$) lighter than the top quark
and the other squarks.

H1 and ZEUS  reported an event excess relative to the SM expectation at the
large Bjorken parameter $x$ and high four-momentum transfer squared $Q^2$ in
the deep inelastic scattering (DIS),   $e^+p  \to  e^+X$ \cite {h1,zeus}.
Although the excess became less significant recently with increase of
statistics,  this fact implies that HERA is a powerful machine to open the
door  beyond the SM  as leptoquark models and SUSY models with R-parity
breaking (RB) interactions \cite{barger}.

In this letter,  we estimate production cross sections  of the s-channel
resonance of a single stop through RB interactions,   $ep \to
\widetilde{t_1}X\to eX$ with longitudinally polarized  $e^-$ and $e^+$
beams in the framework of the Minimal SUSY Standard Model(MSSM).    Single
stop production  $eq \to \stl \to eq$ in the neutral current processes gives
the sharp peak in the $x$ distribution of the scattered  $e^-$ and $e^+$ ,
and this peak is enhanced in the high $Q^2$ region.  We show that right
handed  $e^+$  beams are more efficient to produce the stop than unpolarized
ones or  $e^-$  beams in the model. HERA will soon be upgraded,
and experiments using polarized  $e^-$ or $e^+$  beams with  high luminosity
will start\cite{kenlong}.  Then we are able to search for the stop with high
statistics in the wide mass range even for  smaller values of the coupling
of  the RB interactions than before.

\section{\it Theoretical models }

We are  based on the minimal MSSM with  RB interactions
\begin{equation}
L=\lambda'_{1jk} ({\widetilde{u_{jL}}} {\overline{d_k}} P_L e
- {\overline{\widetilde{d_{kR}}}} \overline{e^c} P_L u_j) + h.c,
\label{sqRb}
\end{equation}
where $i, j, k$ are generation indices and $P_{L,R}$ are left and right
handed chiral projection operators, respectively.   The Lagrangian Eq.
(\ref{sqRb}) will be most suitable for the $ep$ collider experiments at HERA
because the squarks  will be produced in the $s$-channel in $e^{\pm}$-$q$
sub-processes.
\begin{eqnarray}
&& e^+ (e^-)  + d_k ({\overline{d_k}} )\to {\widetilde{u_{jL}}}
({\overline{\widetilde{u_{jL}}}}) \to e^+(e^-) + d_k  ({\overline{d_k}} )\\
&& e^+  (e^-) + {\overline{u_j}} (u_j ) \to
{\overline{\widetilde{d_{kR}}}}({\widetilde{d_{kR}}})\to e^+ (e^-)+
{\overline{u_j}} (u_j).
\end{eqnarray}

Note that the squark ${\widetilde{u_{jL}}} $or ${\widetilde{d_{kR}}}$ cannot
couple to any neutrinos via the $R$-breaking interactions.
This is a unique property of the squark scenarios which could be useful
for us to distinguish the squarks from some leptoquarks.

Here we pay attention to  the stop production via $\lambda'_{131} \neq 0$ RB
interaction because it could be the lightest squark in the MSSM.  The stops

(${\widetilde{t_L}}$, ${\widetilde{t_R}}$) are naturally mixed each
other \cite{hikasa,stop} due to a large top quark mass \cite{top}
and the mass eigenstates ($\stl$, $\sth$)
are parametrized by a mixing angle $\tht$,
\begin{eqnarray}
&& \stl  = {\widetilde{t_L}}\cos\tht - {\widetilde{t_R}}\sin\tht ,\\
&& \sth = {\widetilde{t_L}}\sin\tht + {\widetilde{t_R}}\cos\tht.
\end{eqnarray}
In this case the interaction Lagrangian (\ref{sqRb}) for the lighter stops
$\stl$ is written
by
\begin{equation}
L=\lambda'_{131} \cos\tht (\stl {\overline{d}} P_L e
                                                 +\stl ^*{\overline{e}} P_R
d ),
\label{stRb}
\end{equation}
which generates the s-channel stop $\stl$ production in the neutral
current(NC) process\cite{stoprb}:
\begin{equation}
e^{\pm}p \to \widetilde{t_1}X\to e^{\pm}qX
\label{ept1x}
\end{equation}
and the relevant Feynman diagrams are depicted in Fig.1. In the calculation,
the contributions of the s-channel  and the u-channel exchange processes of
the stop   are estimated separately, where interferences with amplitudes of
the SM shown in Fig.1 are  taken into account.
We calculate the inclusive differential cross section for the NC processes
$e^{\pm}p \to  e^{\pm}qX$ with longitudinally polarized $e^{\pm}_{L,R}$
beams:

\begin{equation}
 \frac{d\sigma}{dxdQ^2}[e^{\pm}_{L,R}] = \frac{2 \pi \alpha^2}{x^2 s^2}
\sum_{q}[q(x,Q^2)  \sum_{i}  T_i(e^{\pm}_{L,R} q) +    \overline{q}(x,Q^2)
\sum_{i} T_i(e^{\pm}_{L,R} \overline{q})] ,
\label{dxdq2}
\end{equation}

where  $q(x,Q^2)$'s  are the quark distribution functions  in the proton.
The index $i$ represents each diagram in Fig.1 and their interferences. The
analytic expression for the coefficients $T_i(e^{\pm}_{L,R}  q)$is
explicitly given in \cite{stopt}.

The cross section depends sensitively on the decay width of the stop.   In
this calculation, we assume BR( $ \widetilde{t_1} \to ed$  ) $\simeq $ 100
\% , i.e.,
\begin{equation}
\Gamma_{\widetilde{t_1}}= \frac{\alpha}{4}F_{RB}(\widetilde{t_1} )
m_{\widetilde{t_1}},
\label{gam}
\end{equation}
where the coupling strength $F_{RB}(\widetilde{t_1} ) $ is defined as
\begin{equation}
F_{RB}(\widetilde{t_1} ) = \frac{{\lambda'_{131}}^2 { \cos}^2
\tht}{4\pi\alpha}.
\label{frb}
\end{equation}

We expect a clear signal of the stop as a sharp peak in the Bjorken
parameter $x$ distribution at the position of  $x=\mstl^2/s$.  We  point out
from(\ref{stRb})  that the left-handed polarized electron ( $e^{-}_{L}$ )
and the right-handed polarized positron  ($e^{+}_{R}$ )  beams could be
advantageous to produce the stop. Furthermore  the $e^+$ beam is more
efficient than the $e^-$ one to separate the stop signal from the SM
background, which  can be easily understood from the structure of the
coupling (\ref{stRb}).   While the $e^-$ collides only with sea
$\overline{d}$-quarks in the proton , the $e^+$ collides with valence
$d$-quarks to produce the stop.   From the difference of the structure
function of the proton  we can expect   $\sigma (e^+d\to \widetilde{t_1})
 >\sigma (e^-\bar{d} \to \bar{\widetilde{t_1}})$.
Throughout the present work, we
use the CTEQ4M for the parton distribution function\cite{cteq4m}.

\section{\it Cross sections of the stop production}
We have calculated  the differential cross sections  eq.(\ref{dxdq2})  using
the program package BASES \cite{kawabata}, where the energies $E_e$=30 GeV
and $E_p$ =820 GeV ($\sqrt{s}$=314 GeV) are assumed.  Event generation  of
$e^{\pm}p \to  e^{\pm}qX$ has been carried out  by using
SPRING\cite{kawabata} following the numerical results of cross sections
obtained by BASES.   In the previous work, we found that the lower $Q^2$
cuts are efficient to supress the SM background since the s-channel stop
contribution is independent of $Q^2$ \cite{stopt}. In what follows we impose
the condition  $Q^{2}>$  10$^4$ (GeV/c)$^2$ under $Q_{max}^2$ = 98000
(GeV/c)$^2$.

 Fig.2 shows  cross sections of the stop production together with those of
the SM  against the degree of polarization of $e^-$ and $e^+$ beams.
Suggested by H1 and ZEUS  reports \cite {h1,zeus}, we assume  here  $\mstl$
=200 GeV and  $\lambda'_{131}$=0.01, 0.02, 0.03.  These values of
$\lambda'_{131}$ are those below the present upper bound for this value of
$\mstl$ \cite{matsu,apv}.   To compare behavior of the stop production with
the SM processes,  curves obtained by  the $\lambda'_{131}$=0.1 are added
to both $e^-$ and $e^+$ cases.  Dependence of the cross section on the
polarization is stronger in the stop production than in the SM processes.
For example, cross sections of the stop production with fully polarized
$e^-_L$ and $e^+_R$ beams are two times as much as  unpolarized ones while
those of the SM  increase by only  $\sim$30\%.  One can see from this figure
that  righted-handed $e^+$ beams are efficient to produce the stop in our
model.

In order to study production processes of the stop through the s- and the
u-channel ( Fig.1), we calculate separately these contribution for
unpolarized beams.  Fig.3 shows cross sections of the stop production of the
s- and the u-channel processes together with the SM background as a function
of the mass of the stop  $\mstl$.  Here   we assume   that
$\lambda'_{131}$=0.01,
0.03  and other conditions are the same as Fig.2. If  the mass of the stop
is larger than the center-of-mass energy of the $ep$ system, $\sqrt{s}$, a
peak in $x$ distribution disappears. However  contribution of the stop
still remains in the whole region of $x$ distributions due to tails  of the
s-channel resonance and the u-channel process.   From Fig.3  we can see that
contribution of the stop production from the u-channel process  is of the
order of a few   fb.  It is larger  for  $e^-$ beams than for  $e^+$ beams.
This can also be    understood from the structure of the RB interaction.
While the $e^-$ collides with valence  $d$-quarks in the proton  , the $e^+$
collides only with sea $\overline{d}$-quarks  in u-channel diagram.   On the
other hand contribution of the stop from the  tails of the s-channel
resonance   is still   larger  than  the  u-channel contribution in both
$e^-$ and $e^+$ cases. This is due to the broad distribution for the large stop
mass.

 Fig.4 shows  an example of $x$ distributions of scattered   $e^+$  for
fully polarized  righted-handed beams and unpolarized beams under the condition
of the integrated luminosity of 1 fb$^{-1}$, where $\mstl$=200 GeV,
$\lambda'_{131}$=0.015  are assumed as  typical values for our interest.
Peaks at  $x$=0.41 due to the stop over the smooth behavior of the SM
events are more enhanced in the case of polarized $e^+$ beams  than
unpolarized ones.  This implies a possibility to search for the stop at the
level of  $\lambda'_{131}$=0.015 for $e^+$ beams.

The reach in  $\lambda'_{131}$  for the wide range of masses of the stop
160-400 GeV is studied with an integrated luminosity  of 1 fb$^{-1}$. Lower
limits on 95\%confidence-level exclusion for    $\lambda'_{131}$ versus the
mass of the stop  is shown in Fig.5 for  fully polarized   $e^+_R$ beams and
unpolarized ones.  We have  imposed  here
the condition of $S/\sqrt{B}$=1.96 where  $S$ and $B$ denote the number of
stop events and that of the SM background events respectively. Excluded
region $\mstl\le$240 GeV given by the Tevatron  experiments  is also shown in
Fig.5 \cite{cdf}.   We see from this figure that we can search for the stop
having the mass  in the range  of 240-280 GeV at the level of
$\lambda'_{131}>$ 0.01-0.05 with fully polarized $e^+_R$ beams. The upgraded
HERA would meet our expectation.   We can set  the  95\%  confidence-level
exclusion limit in this parameter region if no signal of the stop is
observed.

It is important to identify the stop above the backgrounds due to the SM
process by using kinematical variables characteristics to the stop
resonance.  Differential cross sections of the s-channel stop production
are not sensitive to $Q^2$ in contrast to the strong decrease for  the SM
when  $Q^2$ increases.  It is possible to estimate the contribution of the stop
production by using  $y(=Q^2/sx)$ distributions of $e^+$  for  $e^+ p \to
e^+ X $. Fig.6 shows $y$ distributions of the $e^+$ at the fixed $x=x_{peak}$
for  the stop production together with  the SM.  Here we  assume
$\lambda'_{131}$=0.04 and $\mstl$=250 GeV, which corresponds to $x_{peak}$ =
0.634.
We see from this figure that  possible events in the
$y$  distribution at large  $y$ could be the signature of the stop production.

\section{\it Concluding remarks}
We have investigated the production rate of the stop with polarized electron
and
positron beams at HERA  in the framework of the MSSM with  RB
interactions.   Fully polarized $e^-_L$ or  $e^+_R$ produces the stop two
times as much as unpolarized beams, while they increase the SM backgrounds
by $\sim$30\% in comparison with unpolarized ones. The $e^+$ beams are more
efficient to produce the stop than $e^-$ beams because of the coupling of
$e^+d$ and $e^- \overline{d}$ for the stop in the model.    We can search
for the stop using  $e^+_R$ beams for masses in the range of 240-280 GeV at the
level of   $\lambda'_{131}>$ 0.01-0.05.  Experimentally,  these parameter
ranges
have not been  covered yet.   If no signal of the stop is observed, we can
set  95\%
confidence-level exclusion limit in this parameter region.  We see that
the high $y$ region is effective to identify the stop signal.   The present
limits
would be influenced by the limited degree of polarization and detection
efficiency.

It is pointed out by T.Plehn et al.\cite{kfact} that the contribution of
next-to-leading order (NLO) QCD corrections are important  in the cross
sections
for the formation of scalar resonances, leptoquarks or squarks in $e^\pm p$
collisions at HERA. They showed that the $K$-factors increase
the production cross sections by up to 30\%  depending mildly on the mass of
the resonances, if the target quarks are valence quarks.   In our present
study,  the sea $( \overline{d})$ quark scattering and the off-resonance (s-
and u-channels)
contributions play an important role. It is needed to estimate the NLO
corrections to the whole scattering processes  $e^\pm p \to e^\pm X $.

\begin{flushleft}
{\Large{\bf Acknowledgements}}
\end{flushleft}
This work  was supported in part by the Grant-in-Aid for Scientific Research
from the Ministry of Education, Science and Culture of Japan, No. 10440080.




\begin{figure}[p]
\centerline{\epsfig{figure=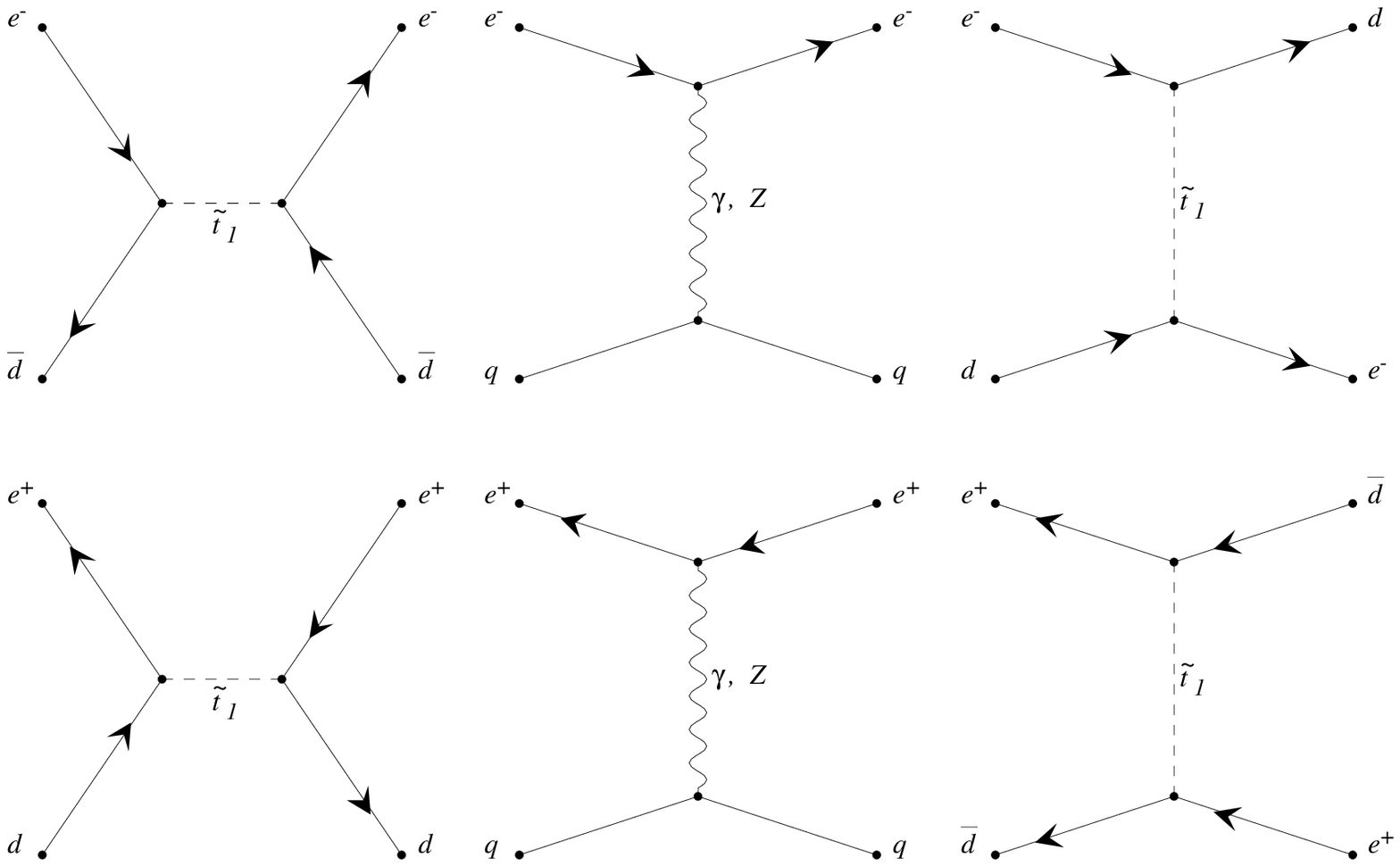,height=10cm,angle=0}}
 \caption{ Feynman diagrams for sub-processes $ e^{\pm}q \to e^{\pm}q $.
 \label{fig1}}
\end{figure}

\begin{figure}[p]
\vspace{10cm}
\centerline{\epsfig{figure=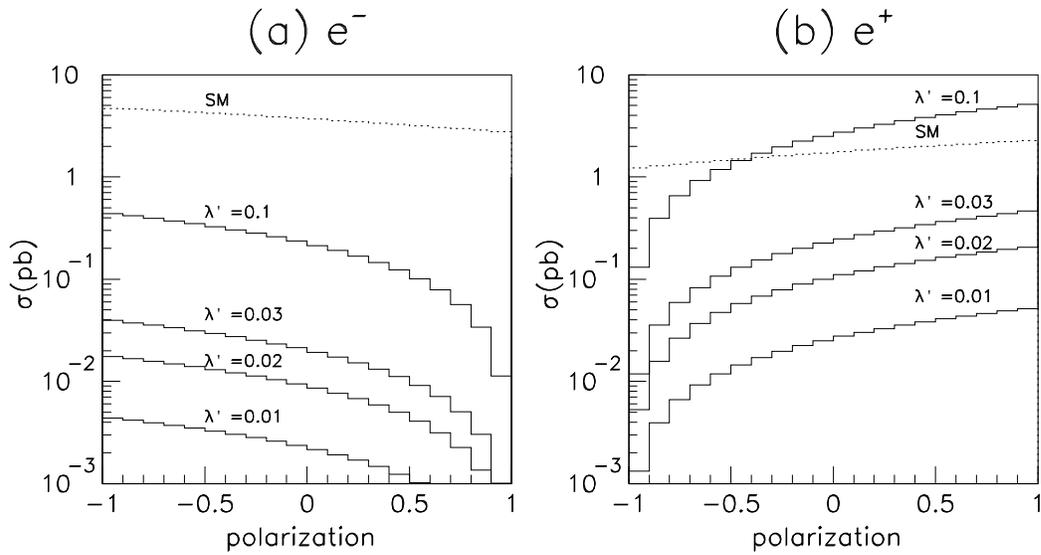,height=12cm,angle=0}}
\vspace{-10cm}
 \caption{ Cross sections of the stop production and the SM process of (a)
$e^-$ and
(b) $e^+$ beams as a function of polarization in $ep$ collisions.
Polarization of $-$1,0,1 corresponds to fully polarized left-handed beam,
unpolarized beam, and fully polarized right-handed beam respectively. Solid
lines  and dotted lines  show the cross sections of stop production and
that of the SM process, respectively.  $ Q^2 >$ 10$^4$
(GeV/c)$^2$, $\mstl$ = 200GeV, $\lambda'_{131}$ = 0.01,0.02,0.03,  0.1 and
$\theta_t$ =0.0  are assumed.
 \label{fig2}}
\end{figure}

\begin{figure}[p]
\vspace{10cm}
\centerline{\epsfig{figure=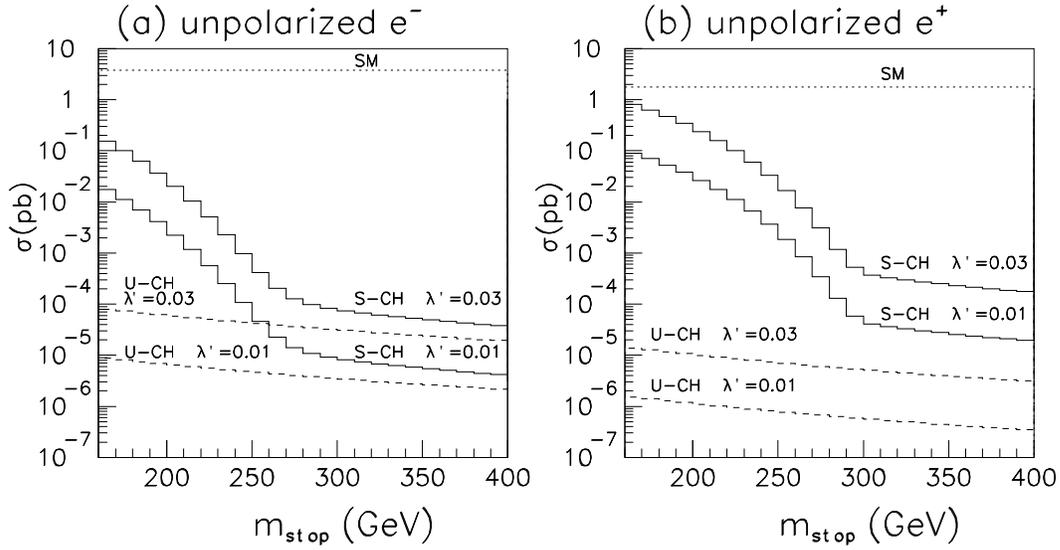,height=12cm,angle=0}}
\vspace{-10cm}
 \caption{ Cross sections  of the stop production due to  s-channel process
(solid
lines) , u-channel process  (dashed lines)  and the SM background (dotted
lines) as a function of the mass of the stop $\mstl$  for (a)  $e^-$
unpolarized beams   and (b) $e^+$ unpolarized beams.  $ Q^2
>$ 10$^4$ (GeV/c)$^2$, $\lambda'_{131}$ = 0.01, 0.03 and  $\theta_t$ =0.0  are
assumed in both cases.
 \label{fig3}}
\end{figure}

\begin{figure}[p]
\vspace{10cm}
\centerline{\epsfig{figure=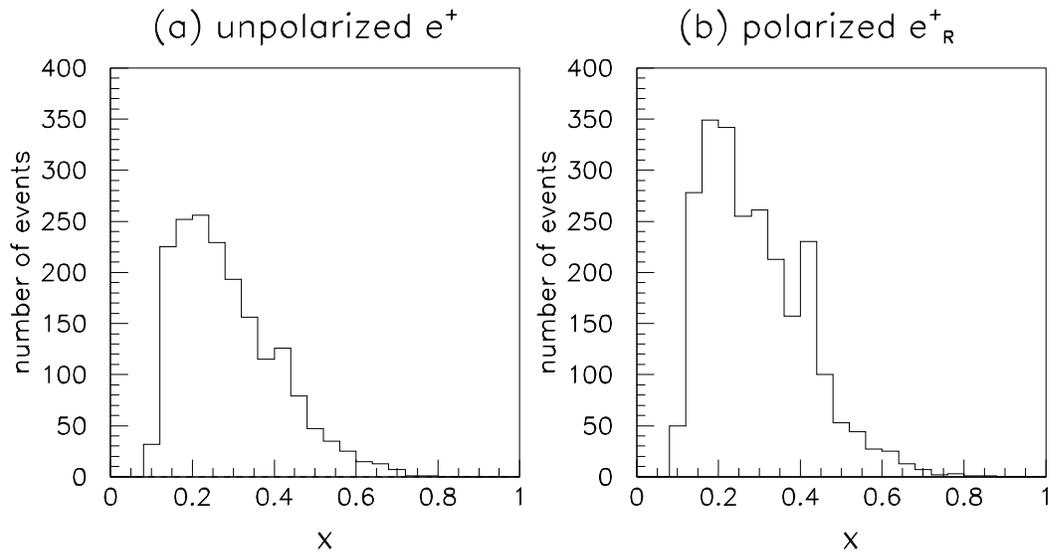,height=12cm,angle=0}}
\vspace{-10cm}
 \caption{ $x$ distributions of $e^+$ of the processes $ e^+p \to e^+X $
with  (a)
unpolarized $e^+$ beams and  (b) fully polarized $e^+_R$ beams.  $ Q^2 >$
10$^4$  (GeV/c)$^2$ , $\mstl$ = 200GeV,  $\lambda'_{131}$ =
0.015 and  $\theta_t$ =0.0 are assumed  with ${\cal L}$ = 1 fb$^{-1}$.
 \label{fig4}}
\end{figure}


\begin{figure}[p]
\vspace{10cm}
\begin{center}
\centerline{\epsfig{figure=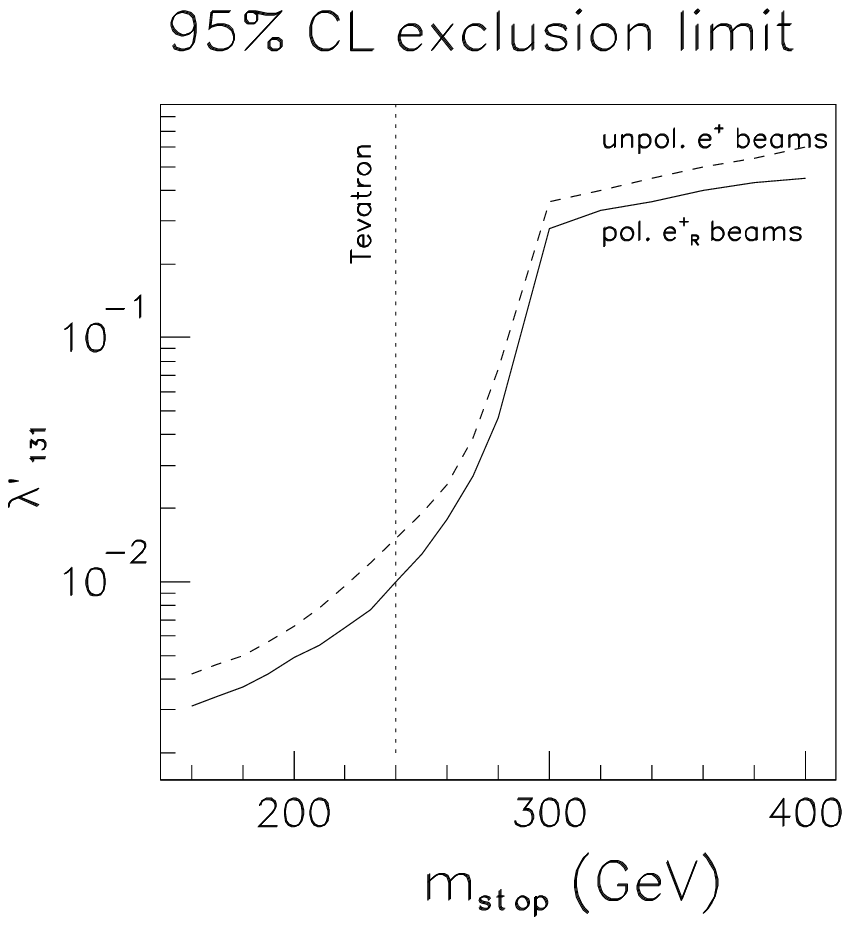,height=15cm,angle=0}}
\end{center}
\vspace{-15cm}
 \caption{ 95$\%$ confidence-level exclusion limit $(S/\sqrt{B}=1.96)$ for
$\lambda'_{131}$ versus stop mass $\mstl$ in run of HERA  with an
integrated luminosity  of 1fb$^{-1}$ for fully polarized $e^+_R$
beams (solid line) and unpolarized $e^+$ beams(dashed line).
Upper region of each curve is excluded. A dotted
vertical line is a lower bound of $\mstl$ of Tevatron\cite{cdf}.
 \label{fig5}}
\end{figure}

\begin{figure}[p]
\vspace{10cm}
\centerline{\epsfig{figure=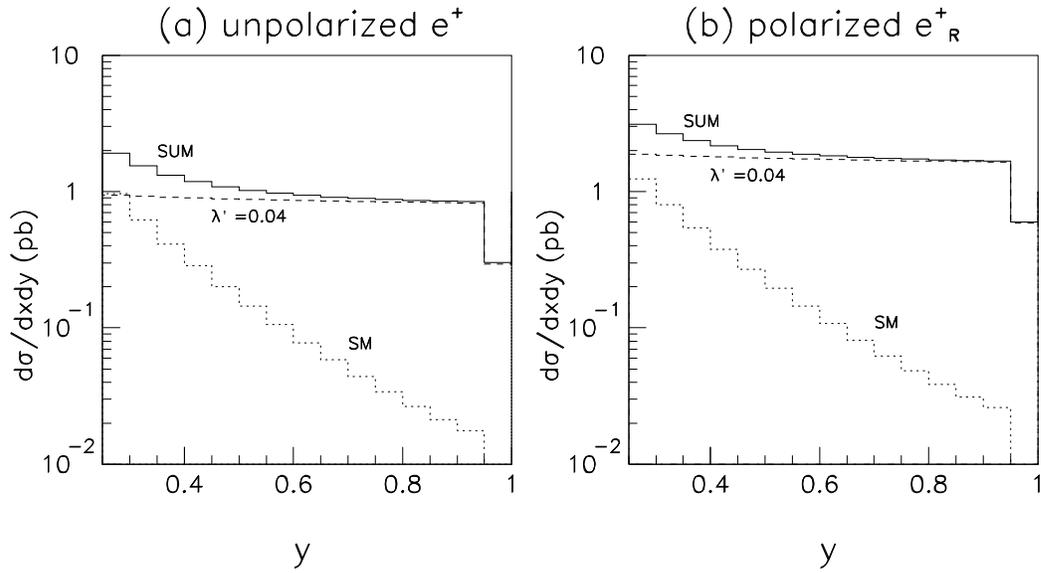,height=12cm,angle=0}}
\vspace{-10cm}
 \caption{ $y$-distributions  of $e^+$ of the process  $e^+ p \to e^+ X $ for
(a) unpolarized $e^+$  beam and (b) fully polarized $e^+_R$ beams,
where $x$
is within the interval  0.634$\pm$0.02 which corresponds to $\mst$ =
250$\pm$4GeV.  $Q^2 >$ 10$^4$ (GeV/c)$^2$,
$\lambda'_{131}$
= 0.04  and $\theta_t$ =0.0  are assumed.  Dashed, dotted and solid lines
show the distributions for the process of  $\stl$  production, the SM
process and sum of them, respectively.
 \label{fig6}}
\end{figure}


\vfill\eject

\end{document}